\documentclass{PoS}

\def\fun#1#2{\lower3.6pt\vbox{\baselineskip0pt\lineskip.9pt
\ialign{$\mathsurround=0pt#1\hfil##\hfil$\crcr#2\crcr\sim\crcr}}}

\newcommand{\beq}{\begin{equation}}
\newcommand{\eeq}{\end{equation}}
\newcommand{\bea}{\begin{eqnarray}}
\newcommand{\eea}{\end{eqnarray}}

\DeclareSymbolFont{boldletters}{OML}{cmm} {b}{it}
\DeclareSymbolFontAlphabet{\mathbit}{boldletters}
\DeclareMathSymbol{\alpha}{\mathalpha}{letters}{"0B}
\DeclareMathSymbol{\beta}{\mathalpha}{letters}{"0C}
\DeclareMathSymbol{\gamma}{\mathalpha}{letters}{"0D}
\DeclareMathSymbol{\delta}{\mathalpha}{letters}{"0E}
\DeclareMathSymbol{\epsilon}{\mathalpha}{letters}{"0F}
\DeclareMathSymbol{\zeta}{\mathalpha}{letters}{"10}
\DeclareMathSymbol{\eta}{\mathalpha}{letters}{"11}
\DeclareMathSymbol{\theta}{\mathalpha}{letters}{"12}
\DeclareMathSymbol{\iota}{\mathalpha}{letters}{"13}
\DeclareMathSymbol{\kappa}{\mathalpha}{letters}{"14}
\DeclareMathSymbol{\lambda}{\mathalpha}{letters}{"15}
\DeclareMathSymbol{\mu}{\mathalpha}{letters}{"16}
\DeclareMathSymbol{\nu}{\mathalpha}{letters}{"17}
\DeclareMathSymbol{\xi}{\mathalpha}{letters}{"18}
\DeclareMathSymbol{\pi}{\mathalpha}{letters}{"19}
\DeclareMathSymbol{\rho}{\mathalpha}{letters}{"1A}
\DeclareMathSymbol{\sigma}{\mathalpha}{letters}{"1B}
\DeclareMathSymbol{\tau}{\mathalpha}{letters}{"1C}
\DeclareMathSymbol{\upsilon}{\mathalpha}{letters}{"1D}
\DeclareMathSymbol{\phi}{\mathalpha}{letters}{"1E}
\DeclareMathSymbol{\chi}{\mathalpha}{letters}{"1F}
\DeclareMathSymbol{\psi}{\mathalpha}{letters}{"20}
\DeclareMathSymbol{\omega}{\mathalpha}{letters}{"21}
\DeclareMathSymbol{\varepsilon}{\mathalpha}{letters}{"22}
\DeclareMathSymbol{\vartheta}{\mathalpha}{letters}{"23}
\DeclareMathSymbol{\varpi}{\mathalpha}{letters}{"24}
\DeclareMathSymbol{\varrho}{\mathalpha}{letters}{"25}
\DeclareMathSymbol{\varsigma}{\mathalpha}{letters}{"26}
\DeclareMathSymbol{\varphi}{\mathalpha}{letters}{"27}
\DeclareMathSymbol{\Gamma}{\mathalpha}{letters}{"00}
\DeclareMathSymbol{\Delta}{\mathalpha}{letters}{"01}
\DeclareMathSymbol{\Theta}{\mathalpha}{letters}{"02}
\DeclareMathSymbol{\Lambda}{\mathalpha}{letters}{"03}
\DeclareMathSymbol{\Xi}{\mathalpha}{letters}{"04}
\DeclareMathSymbol{\Pi}{\mathalpha}{letters}{"05}
\DeclareMathSymbol{\Sigma}{\mathalpha}{letters}{"06}
\DeclareMathSymbol{\Upsilon}{\mathalpha}{letters}{"07}
\DeclareMathSymbol{\Phi}{\mathalpha}{letters}{"08}
\DeclareMathSymbol{\Psi}{\mathalpha}{letters}{"09}
\DeclareMathSymbol{\Omega}{\mathalpha}{letters}{"0A}

\title{Temperature dependence of meson screening masses; a comparison of effective model with lattice QCD}

\ShortTitle{Temperature dependence of meson screening masses; a comparison of effective model with lattice QCD}

\author{\speaker{Masahiro Ishii}%
        \\
        Department of Physics, Graduate School of Sciences, Kyushu
        University, Fukuoka 812-8581, Japan\\
        E-mail: \email{ishii@phys.kyushu-u.ac.jp}}
\author{Takahiro Sasaki%
        \\
        Department of Physics, The University of Tokyo,
        7-3-1 Hongo, Bunkyo-ku, Tokyo 113-0033, Japan and
        Department of Physics, Graduate School of Sciences, Kyushu
        University, Fukuoka 812-8581, Japan\\
        E-mail: \email{t-sasaki@nt.phys.s.u-tokyo.ac.jp }}

\author{Kouji Kashiwa%
        \\
        Yukawa Institute for Theoretical Physics, Kyoto University, 
        Kitashirakawa Oiwakecho, Sakyo-ku, Kyoto 606-8502, Japan and 
        RIKEN/BNL, Brookhaven, National Laboratory, Upton, New York
        11973, USA\\
        E-mail: \email{kouji.kashiwa@yukawa.kyoto-u.ac.jp }}

\author{Hiroaki Kouno%
        \\
        Department of Physics, Saga University,
             Saga 840-8502, Japan\\
        E-mail: \email{kounoh@cc.saga-u.ac.jp }}

\author{Masanobu Yahiro%
        \\
        Department of Physics, Graduate School of Sciences, Kyushu
        University, Fukuoka 812-8581, Japan\\
        E-mail: \email{yahiro@phys.kyushu-u.ac.jp}}

\abstract{Temperature dependence of pion and sigma-meson screening
masses is evaluated by the Polyakov-loop extended Nambu--Jona-Lasinio model 
with the entanglement vertex (EPNJL model). 
We propose a practical way of calculating meson screening masses 
in the NJL-type effective models. 
The method based on the Pauli-Villars regularization 
solves the well-known difficulty that the evaluation 
of screening masses is not easy in the NJL-type effective models. 
The method is applied to analyze temperature dependence of 
pion screening masses calculated 
with state-of-the-art lattice simulations with success in reproducing 
the lattice QCD results. We predict the temperature 
dependence of pole mass by using EPNJL model.}

\FullConference{The 32nd International Symposium on Lattice Field Theory\\
                 23-28 June, 2014\\
                 Columbia University New York, NY}

\begin{document}

\section{Introduction}
Meson masses are not only fundamental quantities of hadrons but also 
a key to know properties of quantum chromodynamics (QCD) vacuum. At
finite temperature ($T$), we can define two kinds of meson masses, pole
and screening mass. Meson pole masses are one of the possible
observables in the heavy ion collisions. Screening masses
of light mesons are essential for the range of the nuclear
force. Accordingly, it is necessary to construct the effective model for calculating pole and screening mass simultaneously.   

In lattice QCD(LQCD), meson pole (screening) masses are 
calculated from the exponential decay of temporal (spatial) 
mesonic correlation functions. LQCD simulations are more difficult for pole masses 
than for screening masses, since 
the lattice size is smaller in the time direction than 
in the spatial direction. This situation becomes more serious 
as $T$ increases. 
For this reason, meson screening masses were calculated 
in most of the LQCD simulations. Recently, 
a state-of-the-art calculation was done for meson screening masses 
in a wide range of $T < 800$~MeV~\cite{Cheng:2010fe}

Constructing the effective model is an approach complementary to 
 the first-principle LQCD simulation. In contrast to LQCD simulations, meson pole masses 
are extensively investigated at finite $T$  by the  
Nambu--Jona-Lasinio (NJL) model~\cite{Kunihiro,Florkowski}, the Polyakov-loop
extended Nambu--Jona-Lasinio (PNJL) model~\cite{Hansen}. However, only a few trials were made so far for 
the evaluation of meson screening masses 
$M_{\xi,{\rm scr}}$~\cite{Kunihiro,Florkowski}; here $\xi$ means 
a species of mesons. 
The model calculations have essentially two problems. 
One problem is that the NJL-type models are nonrenormalizable and hence 
the regularization is needed in the model calculations. 
The regularization commonly used is the three-dimensional momentum cutoff. 
The momentum cutoff breaks Lorentz and translational invariance, thereby the 
spatial correlation function $\eta_{\xi\xi}(r)$ 
has an unphysical oscillation~\cite{Florkowski}. 
This makes the determination of $M_{\xi,{\rm scr}}$ quite difficult, since 
$M_{\xi,{\rm scr}}$ 
is defined from the exponential decay of $\eta_{\xi\xi}(r)$ 
at large distance ($r$):
\begin{equation}
M_{\xi,{\rm scr}}=-\lim_{r\rightarrow \infty}\frac{d \ln{\eta_{\xi\xi}(r)}}{dr}.
\label{scr-mass}
\end{equation}

\begin{figure}[htbp]
\begin{center}
  \includegraphics[width=0.6\textwidth]{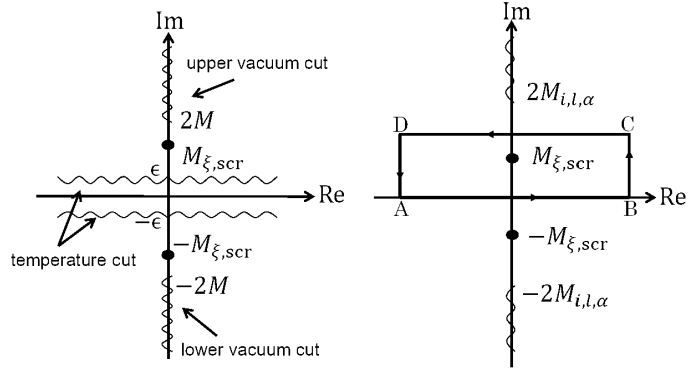}
\end{center}
\caption{Singularities of $\chi_{\xi\xi}(0,\tilde{q}^2)$ 
in the complex-$\tilde{q}$ plane 
based on the previous formulation~\cite{Florkowski} (left)  
and the present formulation 
(right). Cuts are denoted by the wavy lines and poles by the points. 
}
\label{Fig-sing}
\end{figure}

Another problem is the feasibility of numerical calculations. 
In the model approach,  $\eta_{\xi\xi}(r)$ is first obtained 
in the momentum ($\vec{q}$) representation 
$\chi_{\xi\xi}(0,\vec{q}^2)$. 
In the Fourier transformation to the coordinate representation ($r=|\vec{x}|$),
\begin{equation}
\eta_{\xi\xi}(r)
=\int\frac{d^3 q}{(2\pi)^3}\chi_{\xi\xi}(0,\vec{q}^2)e^{i\vec{q}\cdot\vec{x}}
=\frac{1}{4\pi^{2}ir}\int^{\infty}_{-\infty}d\tilde{q}\hspace{1ex}\tilde{q}\chi_{\xi\xi}(0,\tilde{q}^2)e^{i\tilde{q}r}\hspace{1ex}.
\label{chi_r}
\end{equation}
The integrand is slowly damping and highly oscillating particularly 
at large $r$ where $M_{\xi,{\rm scr}}$ is defined.
This requires heavy numerical calculations. 
It was then proposed that the contour integral was made 
in the complex-$\tilde{q}$ plane~\cite{Florkowski}. 
However, the contour integral is 
still hard to do because of the presence of the temperature cuts 
in the vicinity of the real axis~\cite{Florkowski}; see 
the left panel of Fig.~\ref{Fig-sing}, where note that 
$\epsilon$ is an infinitesimal quantity. 

In this talk, we propose a new formalism for calculating screening mass
and discuss the possibility of the prediction for pole mass from
screening mass by using effective model. This talk is based on the paper~\cite{Ishii}.

\section{Formalism}
\label{Formalism}
The Lagrangian density of the two-flavor EPNJL model~\cite{Sakai_EPNJL} is 
defined as 
\begin{equation}
 {\cal L}  
= {\bar q}(i \gamma_\nu D^\nu -m_0)q  + G_{\rm s}(\Phi)[({\bar q}q )^2 
  +({\bar q }i\gamma_5 {\vec \tau}q )^2]
 -{\cal U}(\Phi [A],{\bar \Phi} [A],T) 
\label{L}
\end{equation}
with the quark field $q$, the current quark mass $m_0$ and 
the isospin matrix ${\vec \tau}$. 
The coupling constant $G_{\rm s}(\Phi)$ of the 
four-quark interaction depends on the Polyakov loop $\Phi$ as 
\begin{equation}
G_{\rm s}(\Phi)=G_{\rm s}\left[1-\alpha_1\Phi{\bar \Phi} -\alpha_2\left(\Phi^3 + {\bar \Phi}^{3}\right)\right],
\label{EPNJL}
\end{equation}
where 
$D^\nu=\partial^\nu + iA^\nu$ with $A^\nu=\delta^{\nu}_{0}g(A^0)_a{\lambda_a/2}
=-\delta^{\nu}_{0}ig(A_{4})_a{\lambda_a/2}$ for the gauge field $A^\nu_a$, the Gell-Mann matrix $\lambda_a$ and the gauge coupling $g$. 
When $\alpha_1=\alpha_2=0$, the EPNJL model is 
reduced to the PNJL model~\cite{Hansen}.

In the EPNJL model, only the time component of $A_\mu$ is treated as a
homogeneous and static background field, which is governed by the
Polyakov-loop potential~~$\mathcal{U}$. The Polyakov loop $\Phi$ and 
its conjugate ${\bar \Phi}$ are then obtained in the Polyakov gauge by  
\begin{eqnarray}
\Phi &= {1\over{3}}{\rm tr}_{\rm c}(L),
~~~~~{\bar \Phi} ={1\over{3}}{\rm tr}_{\rm c}({L^*})
\label{Polyakov}
\end{eqnarray}
with $L= \exp[i A_4/T]=\exp[i {\rm diag}(A_4^{11},A_4^{22},A_4^{33})/T]$ 
for the classical variables $A_4^{ii}$ 
satisfying that $A_4^{11}+A_4^{22}+A_4^{33}=0$. For zero chemical potential, $\Phi$ equals to ${\bar \Phi}$. Hence it is possible to set $A^{33}_4=0$ and determine the others as 
$A^{22}_4=-A^{11}_4={\rm cos}^{-1}(\frac{3\Phi -1}{2})T$. We use the logarithm-type Polyakov-loop potential $\mathcal{U}$ of
 Ref.~\cite{Rossner}, but refit the parameter $T_0$ to reproduce the chiral
phase transition temperature $T_c$ because the original value of $T_0$
is set to $270$ MeV which is the deconfinement transition temperature in the pure gauge limit.

Making the mean field approximation to (\ref{L}) and the path integral over the quark field, one can get 
the thermodynamic potential (per unit volume) as
\begin{eqnarray}
\Omega= U_{\rm M}+{\cal U}-2 N_{\rm f} \int \frac{d^3 p}{(2\pi)^3}
\Bigl[ 3 E_p 
&+& \frac{1}{\beta}
           \ln~ [1 + 3(\Phi+{\bar \Phi} e^{-\beta E_p}) 
           e^{-\beta E_p}+ e^{-3\beta E_p}] \nonumber\\
&+& \frac{1}{\beta} 
           \ln~ [1 + 3({\bar \Phi}+{\Phi e^{-\beta E_p}}) 
              e^{-\beta E_p}+ e^{-3\beta E_p}]
	      \Bigl]
	      \nonumber\\
\label{PNJL-Omega}
\end{eqnarray}
with $\beta=1/T$, $M=m_0-2G_{\rm
s}(\Phi)\sigma$, $E_p=\sqrt{\vec{p}^2+M^2}$,  and  $U_{\rm M}= G_{\rm
s}(\Phi)\sigma^2$. Here,
$\sigma$ means chiral condensate $\langle\bar{q}q\rangle$. $N_{\rm f}$ is the number of flavors. We determine the mean
field variables ($X=\sigma,\Phi,\bar{\Phi}$) from the stationary
conditions for $\Omega$,
\begin{equation}
\frac{\partial \Omega}{\partial X}=0~.
\end{equation}

Since the momentum integral of (\ref{PNJL-Omega}) diverges, 
we use the Pauli--Villars (PV) regularization~\cite{Florkowski,PV}. 
In the scheme, the integral $I(M,q)$ is regularized as 
\begin{eqnarray}
I^{\rm reg}(M,q)=\sum_{\alpha=0}^2 C_\alpha I(M_\alpha,q) ,
\label{PV}
\end{eqnarray}
where $M_0=M$ and $M_\alpha~(\alpha\ge 1)$ are masses of auxiliary 
particles. The parameters $M_\alpha$ and $C_\alpha$ 
should satisfy the condition  
$\sum_{\alpha=0}^2C_\alpha=\sum_{\alpha=0}^2 C_\alpha M_\alpha^2=0$. We then assume $(C_0,C_1,C_2)=(1,1,-2)$ and $(M_1^2,M_2^2)=(M^2+2\Lambda^2,M^2+\Lambda^2)$. We keep the parameter $\Lambda$ finite 
even after the subtraction (\ref{PV}), since the present model is 
nonrenormalizable. 
The parameters taken are 
$m_0=6.3$ MeV, $G_{\rm s}=5.0$ GeV$^{-2}$ and $\Lambda =0.768$ GeV. 
This parameter set reproduces the pion decay constant $f_{\pi}=93.3$ MeV and 
the pion mass $M_{\pi}=138$ MeV at vacuum. 


We derive the equations for pion and sigma-meson masses, 
following Ref~\cite{Hansen}. We consider currents with the same quantum number 
as pion ($P$) and sigma-meson ($S$),
\begin{eqnarray}
  {J_P}^a(x) = \bar q(x) i \gamma_5 \tau^a q(x)~,~ 
  {J_S}(x)   = \bar q(x) q(x) - \langle\bar q(x) q(x) \rangle .
\end{eqnarray}
The Fourier transform of the mesonic correlation function 
$\eta_{\xi\xi} (x) \equiv \langle 0 | T \left( J_\xi(x) J^{\dagger}_{\xi}(0) \right) | 0 \rangle$ is 
\begin{eqnarray}
\chi_{\xi\xi} (q^2) 
=  i \int d^4x ~e^{i q\cdot x}
 \langle 0 | {\rm T} \left( J_\xi(x) J^{\dagger}_{\xi}(0) \right)  
 | 0 \rangle  ,
\end{eqnarray}
where $\xi=P^a$ for pion and $S$ for sigma meson and ${\rm T}$ stands for 
the time-ordered product. Using the random-phase (ring) approximation, one can obtain
$\chi_{\xi\xi}$ as follows,
\bea
\chi_{\xi\xi} &=& \frac{\Pi_{\xi\xi}}{1 - 2G_{\rm s}(\Phi) \Pi_{\xi\xi}}, 
\label{SD-eq}
\eea
where the one-loop polarization function 
$\Pi_{\xi\xi}$ is explicitly obtained by
\begin{eqnarray}
\Pi_{SS}=2iN_{\rm f}[I_1+I_2-(q^2-4M^2)I_3]~,~ 
\Pi_{PP}=2iN_{\rm f}[I_1+I_2-q^2I_3], 
\end{eqnarray}
with 
\begin{eqnarray}
I_1&=&\int {d^4p\over{(2\pi )^4}}{\rm tr_c}\Bigl[{1\over{p'^2-M^2}}\Bigr],~~
\label{I1}
I_2=\int {d^4p\over{(2\pi )^4}}{\rm tr_c}\Bigl[{1\over{(p'+q)^2-M^2}}\Bigr],
\label{I2}
\\
I_3&=&\int {d^4p\over{(2\pi )^4}}{\rm tr_c}\Bigl[{1\over{\{(p'+q)^2-M^2\}(p'^2-M^2)}}\Bigr] ,
\label{I3}
\end{eqnarray}
Here, $q^2 = q_0^2 - \vec{q}^2$ and $p'=(p_{0}+iA_4,\vec{p})$. ${\rm tr}_{\rm c}$ means the trace of color matrix. For finite $T$, the corresponding equations are 
obtained by the replacement
\begin{eqnarray}
p_0 &\to& i \omega_l = i(2l+1) \pi T~,~ 
\int \frac{d^4p}{(2 \pi)^4} 
\to iT\sum_{l=-\infty}^{\infty} \int \frac{d^3p}{(2 \pi)^3}. 
\label{finte_T_mu}
\end{eqnarray}

The meson pole mass $M_{\xi,{\rm pole}}$ is a pole of 
$\chi_{\xi\xi}(q_0^2,\vec{q}^2)$. Taking the rest frame $q=(q_0,\vec{0})$ 
for convenience, one can get the equation for $M_{\xi,{\rm pole}}$ as 
\begin{eqnarray}
\big[1 - 2G_{\rm s}(\Phi) \Pi_{\xi\xi}(q_0^2,0)\big]\big|_{q_0=M_{\xi,{\rm pole}}}=0.   
\label{mmf}
\end{eqnarray}
The method of calculating meson pole masses is well established 
in the PNJL model~\cite{Hansen}.

The meson screening mass $M_{\xi,{\rm scr}}$ defined with (\ref{scr-mass}) 
is obtained by making the Fourier transform of 
$\chi_{\xi\xi} (0,\tilde{q}^{2})$ as shown in (\ref{chi_r}). 
In the previous formalism~\cite{Florkowski}, however, the procedure 
requires heavy numerical calculations in the $I_3^{\rm reg}$ part, 
as shown below, where $I_3^{\rm reg}$ means a function 
after the PV regularization. 
Taking the $l$ summation before the $p$ integral in (\ref{finte_T_mu}), 
one can describe 
$I_3^{\rm reg}(0,\tilde{q}^{2})$ as 
the sum of the vacuum and temperature parts, 
$I_{3,{\rm vac}}^{\rm reg}$ and $I_{3,{\rm tem}}^{\rm reg}$, defined by 
\begin{eqnarray}
I_{3,{\rm vac}}^{\rm reg}(0,\tilde{q}^{2})
&=&\frac{-iN_c}{16\pi^2}\sum_{\alpha=0}^{2}C_\alpha\left[\ln{M_\alpha^2}+f_{\rm vac}\left(\frac{2M_{\alpha}}{\tilde{q}}\right)\right] ,~~~
\\
I_{3,{\rm tem}}^{\rm reg}(0,\tilde{q}^{2})
&=&\frac{iN_c}{16\pi^2}\sum_{\alpha=0}^{2}C_\alpha\int_0^\infty 
d|\vec{p}|~f_{\rm tem}(|\vec{p}|,\tilde{q})\left[F^{+}(E_{p})+F^{-}(E_{p})\right],~~~~~~
\\
f_{\rm
 vac}(x)&=&\sqrt{1+x^2}\ln{\left(\frac{\sqrt{1+x^2}+1}{\sqrt{1+x^2}-1}\right)}
~,~f_{\rm tem}(|\vec{p}|,\tilde{q})
=\frac{1}{E_p}\frac{|\vec{p}|}{\tilde{q}}\ln{\left(\frac{(\tilde{q}-2|\vec{p}|)^{2}+\epsilon^{2}}{(\tilde{q}+2|\vec{p}|)^2+\epsilon^{2}}\right)},
\hspace{4ex} 
\label{f-tem}
\end{eqnarray}
where $F_{\pm}$ are the Fermi distribution functions.
 $F_{\pm}$ are defined as 
\begin{eqnarray}
F^{\pm}(E_{p})=\frac{1}{N_c}\sum_{i=1}^{N_c}{1\over{e^{(E_{p} \pm 
iA^{ii}_4 )/T}+1}} . 
\end{eqnarray}
In (\ref{f-tem}), the $\epsilon^2$ term is added 
to make the $|\vec{p}|$ integral well defined at $\tilde{q}=\pm2|\vec{p}|$, 
but this requires the limit of $\epsilon \to 0$.

As shown in the left panel of Fig.~\ref{Fig-sing}, 
$f_{\rm vac}({2M_{\alpha}}/{\tilde{q}})$ and 
$f_{\rm tem}(|\vec{p}|,\tilde{q})$ have the vacuum and temperature cuts in 
the complex $\tilde{q}$ plane, respectively. 
In (\ref{chi_r}), the cuts contribute to 
the $\tilde{q}$ integral in addition to 
the pole at $\tilde{q}=iM_{\xi,{\rm scr}}$ defined by
\begin{eqnarray}
\big[1 - 2G_{\rm s}(\Phi) \Pi_{\xi\xi}(0,\tilde{q}^2)\big]\big|_{\tilde{q}=iM_{\xi, {\rm scr}}}=0  .  
\label{meson_screening}
\end{eqnarray}
It is not easy to evaluate the temperature-cut contribution, 
since in (\ref{chi_r}) the integrand is slowly damping and highly oscillating 
with $\tilde{q}$ near the real axis in the complex $\tilde{q}$ plane. 
Furthermore we have to take the limit of $\epsilon \to 0$ finally. In
order to avoid this problem, we integrate about $p$ in
(\ref{finte_T_mu}) before taking Matsubara summation
$\sum_l$. Consequently, we can rewrite $I_3^{\rm reg}$ as an infinite
series of analytic function,
\begin{eqnarray}
I_{3}^{\rm reg}(0,\tilde{q}^{2})
={iT\over{4\pi \tilde{q}}}\sum_{i=1}^{N_c}\sum_{l=-\infty}^\infty
\sum_{\alpha=0}^2
C_{\alpha}  \sin^{-1}{\left({{\tilde{q}\over{2}}\over{\sqrt{{\tilde{q}^2\over{4}}+M_{i,l,\alpha}^2}}}\right)}, 
\label{I_3_final}
\end{eqnarray}
where 
\begin{equation}
M_{i,l,\alpha}(T)=\sqrt{M_{\alpha}^2 + \{(2l+1)\pi T+A_{4}^{ii}\}^2 }. 
\label{KK_mode}
\end{equation}
We have numerically checked 
that the convergence of $l$ summation is quite fast in (\ref{I_3_final}). 
Each term of $I_3^{\rm reg}(0,\tilde{q}^2)$ 
has only two cuts starting from $\pm 2iM_{i,l,\alpha}$ 
on the imaginary axis in the complex $\tilde{q}$ plane. 
The cuts are shown 
in the right panel of Fig.~\ref{Fig-sing}. 
The lowest branch point is $\tilde{q}=2iM_{i=1,l=0,\alpha=0}$. Hence 
$2M_{i=1,l=0,\alpha=0}$ is regarded as ``threshold mass'' 
in the sense that the meson screening-mass spectrum 
becomes continuous above the point. 

If $M_{\xi,{\rm scr}}<2M_{i=1,l=0,\alpha=0}$, 
the pole at $\tilde{q}=iM_{\xi, {\rm scr}}$ 
is well isolated from the cut. Hence one can take the contour 
(A$\to$B$\to$C$\to$D$\to$A) shown 
in the right panel of Fig.~\ref{Fig-sing}. The $\tilde{q}$ integral 
of $\tilde{q}\chi_{\xi\xi}(0,\tilde{q}^2)e^{i\tilde{q}r}$ 
on the real axis in (\ref{chi_r}) is then obtained from the residue 
at the pole and the line integral from point C to point D.
The former behaves as $\exp[-M_{\xi,{\rm scr}} r]/r$ at large $r$ 
and the latter as $\exp[{-2M_{i=1,l=0,\alpha=0}}r]/r$. 
The behavior of $\eta_{\xi\xi}(r)$ at large $r$ is thus 
determined by the pole. 
One can then determine the screening mass from the location of 
the pole in the complex-$\tilde{q}$ plane 
without making the $\tilde{q}$ integral. 
In the high-$T$ limit, the condition tends to $M_{\xi,{\rm scr}}<2 \pi T$.

\section{Numerical Results}


 
The pion screening mass $M_{\pi, {\rm scr}}$ obtained by 
state-of-the-art 2+1 flavor LQCD simulations~\cite{Cheng:2010fe}  
is now analyzed by the present two-flavor EPNJL model simply, 
since pion is composed of $u$ and $d$ quarks. 
This is a quantitative analysis, because the finite lattice-spacing effect is 
not negligible in the simulations. 
The chiral transition temperature is evaluated as 
$T_c=196~{\rm MeV}$ 
in the simulations~\cite{Cheng:2010fe}, although it becomes 
$T_c=154\pm9~{\rm MeV}$ 
in finer 2+1-flavor LQCD simulations~\cite{Bazavov} close 
to the continuum limit. 
Therefore, we rescale the LQCD results of Ref.~\cite{Cheng:2010fe} with
multiplying them by the factor $154/196$ to reproduce 
$T_c=154\pm9~{\rm MeV}$. 
The model parameters, $m_0$ and 
$T_{0}$, are refitted to reproduce the rescaled 2+1 flavor LQCD data, i.e., 
$M_{\pi}=175~{\rm MeV}$ at vacuum and $T_c=154\pm9~{\rm MeV}$; 
the resulting values are $m_0=10.3~{\rm MeV}$ and $T_{0}=156~{\rm MeV}$.
The variation of $m_0$ from the original value $6.3$ to 
$10.3~{\rm MeV}$ little changes $\sigma$ and $\Phi$.

\begin{figure}[htbp]
 \begin{minipage}{0.48\hsize}
  \begin{center}
   \includegraphics[width=55mm]{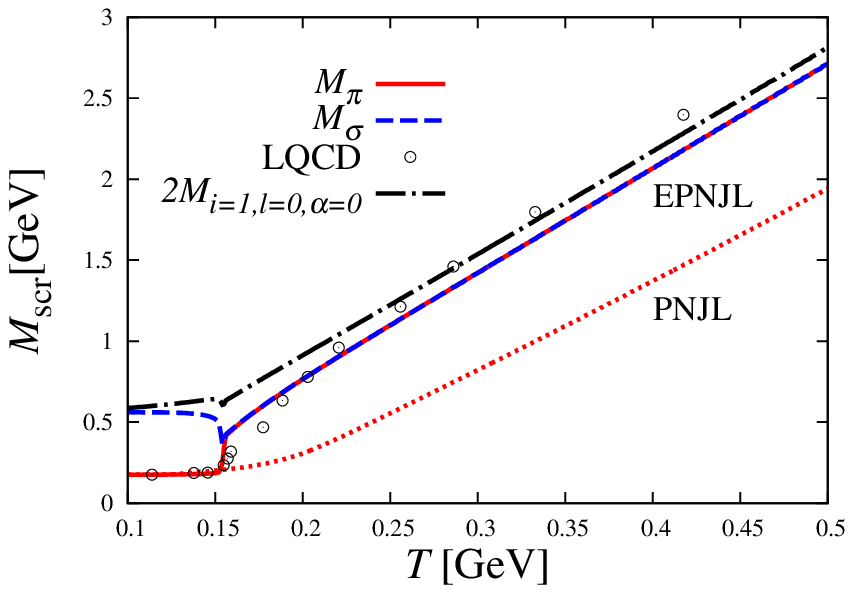}
  \end{center}
\caption{$T$ dependence of pion and sigma-meson screening masses, 
$M_{\pi,{\rm scr}}$ and $M_{\sigma,{\rm scr}}$.}
\label{Fig-screening}
 \end{minipage}
 \begin{minipage}{0.48\hsize}
  \begin{center}
   \includegraphics[width=55mm]{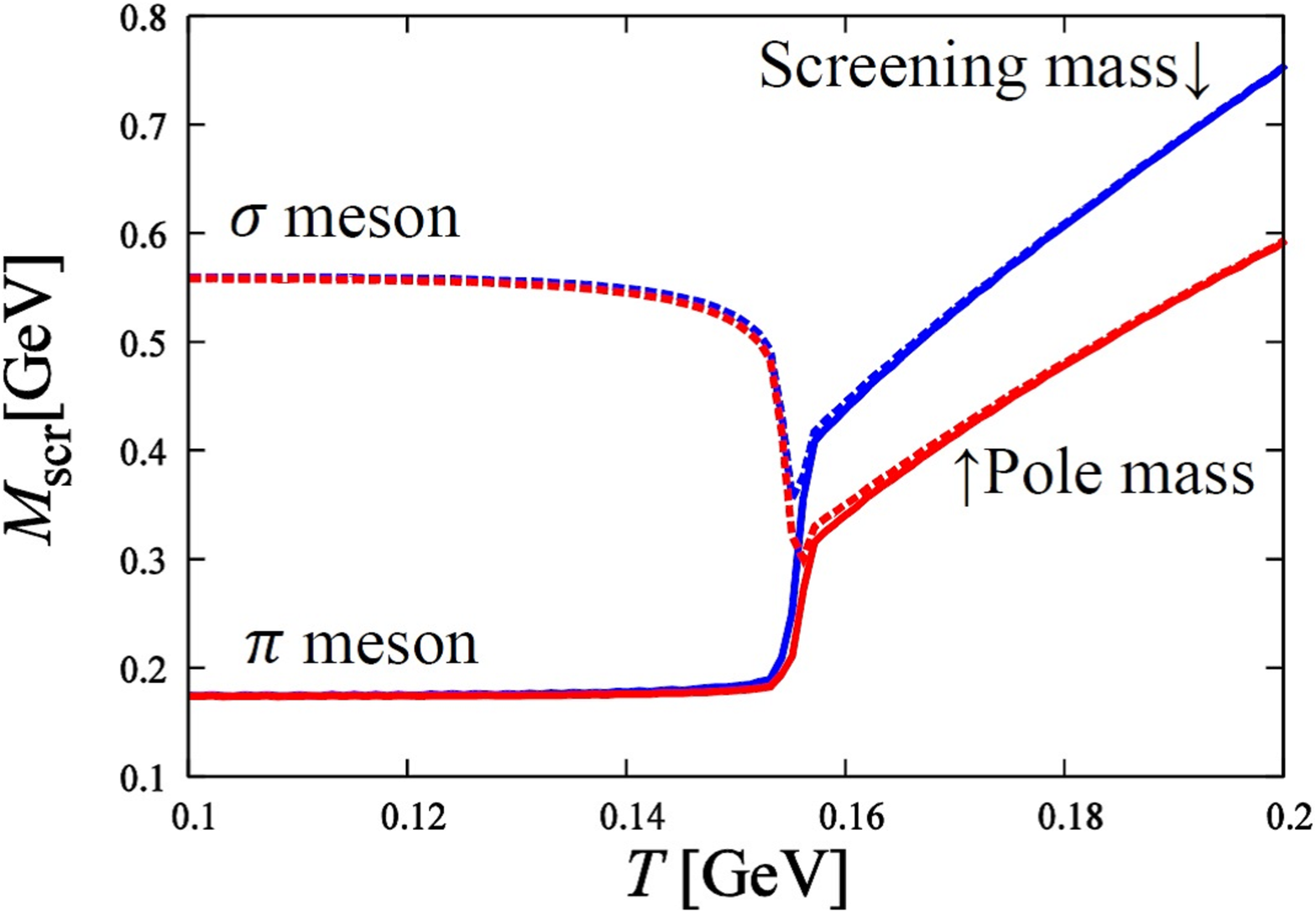}
  \end{center}
  \caption{$T$ dependence of screening and pole mass in pion and sigma-meson.}
  \label{Fig-pole-scr}
 \end{minipage}
\end{figure}

As shown in Fig.~\ref{Fig-screening}, the $M_{\pi, {\rm scr}}$ calculated 
with the EPNJL model (solid line) well reproduces the LQCD result 
(open circles), 
when $\alpha_1=\alpha_2=0.31$.
In the PNJL model with $\alpha_1=\alpha_2=0$, 
the model result (dotted line) largely underestimates the LQCD 
result, indicating that the entanglement is important. 
The dashed line denotes the sigma-meson screening mass 
$M_{\sigma, {\rm scr}}$ obtained by the EPNJL model with 
$\alpha_1=\alpha_2=0.31$. 
The solid and dashed lines are lower than the threshold mass 
$2M_{i=1,l=0,\alpha=0}$ (dot-dashed line). This guarantees 
that the $M_{\pi, {\rm scr}}$ and $M_{\sigma, {\rm scr}}$ determined 
from  the location of the single pole in the complex-$\tilde{q}$ plane 
agree with those from the exponential decay 
of $\eta_{\xi\xi}(r)$ at large $r$. 
The chiral restoration takes place at $T=T_c=154$~MeV, since 
$M_{\pi, {\rm scr}}=M_{\sigma, {\rm scr}}$ there. 
After the restoration, the screening masses 
rapidly approach the threshold mass and finally $2 \pi T$. 
The threshold mass is thus an important concept 
to understand $T$ dependence of screening masses.

Finally, we predict the $T$ dependence of pole mass $M_{\xi,{\rm
pole}}$ for pion and sigma-meson with EPNJL
model (Fig. \ref{Fig-pole-scr}). At low temperature ($T<T_c$), the $T$ dependence
of $M_{\xi,{\rm pole}}$ and $M_{\xi,{\rm scr}}$ are almost same in
the pion and sigma-meson because Lorentz symmetry is preserved
approximately. Around $T_c$, pion
and sigma-meson masses agree with each other and chiral symmetry
restoration takes place at the same temperature $T_c$ for pole and screening
mass. These indicate that at low temperature ($T\lesssim T_c $) we can
predict the $T$ dependence of $M_{\xi,{\rm
pole}}$ from that of $M_{\xi,{\rm scr}}$ simply. Above $T_c$,
however, the
difference between $M_{\xi,{\rm pole}}$ and $M_{\xi,{\rm scr}}$ gets
larger as temperature increases. Therefore, above $T_c$, it is
necessary that we should use the effective model to predict the pole
mass from the lattice QCD results of screening mass.

\section{Summary}

We have proposed a practical way of calculating meson 
screening masses $M_{{\xi,\rm scr}}$ in the NJL-type models. 
This method based on the PV regularization 
solves the well-known difficulty that the evaluation of $M_{{\xi, \rm scr}}$ 
is not easy in the NJL-type effective models. 
In the previous formalism~\cite{Florkowski}, the vacuum and temperature cuts 
appear in the complex-$\tilde{q}$ plane. The contributions 
to the mesonic correlation function are partially canceled 
in the present formalism. The branch point of 
the resultant cut can be regarded as the threshold mass. 
The pion and sigma-meson screening masses 
rapidly approach the threshold mass $2M_{i=1,l=0,\alpha=0}(T)$ 
after the chiral restoration. We propose the prediction for pole mass
from screening mass by using EPNJL model.

\begin{acknowledgments}
The authors thank J. Takahashi for useful discussion. Four of authors
 (T. S., K. K., H. K. and M. Y.) are supported by Grant-in-Aid for
 Scientific Research (No. 23-2790, No. 26-1717, No. 26400279 and No. 26400278) from Japan Society for the Promotion of Science (JSPS)
\end{acknowledgments}

\end{document}